\documentclass[letterpaper,preprintnumbers,eqsecnum,superscriptaddress,aps,nofootinbib,11pt]{revtex4}
\usepackage{graphicx}\usepackage{float}\usepackage{subfigure}\usepackage{titlesec}\usepackage{CJK}
\usepackage{amssymb}\usepackage[centertags]{amsmath}\usepackage{txfonts,epsfig}\usepackage{bm}
\usepackage{color}\usepackage{graphics}\usepackage{multirow,float}\usepackage{ulem}\usepackage{epsfig}
\usepackage{hyperref}\usepackage{setspace,slashed}\usepackage{booktabs}
\usepackage{pdfpages}\usepackage{epstopdf}
\hypersetup{colorlinks=true, citecolor=blue, linkcolor=red, filecolor=black,urlcolor=blue}
\allowdisplaybreaks[4]

\titleformat*{\section}{\bfseries}
\titleformat*{\subsection}{\bfseries}
\titleformat*{\subsubsection}{\itshape\subsubsectionfont}

\begin{document}
\begin{CJK*}{GBK}{song}

\title{Cosmic muon flux measurement and tunnel overburden structure imaging}

\author{Ran HAN}\thanks{Corresponding author}\thanks{hanran@ncepu.edu.cn}
\author{Qian YU}
\affiliation{Science and Technology on Reliability and Environmental Engineering Laboratory, Beijing Institute of Spacecraft Environment Engineering, Beijing 100094, China}

\author{Zhiwei LI}\thanks{Corresponding author} \thanks{zwli@apm.ac.cn}
\affiliation{State Key Laboratory of Geodesy and Earth's Dynamics, Innovation for Precision Measurement Science and Technology, Chinese Academy of Sciences, Wuhan 430077, China}

\author{Jingtai LI}
\author{Yaping CHENG}
\affiliation{Science and Technology on Reliability and Environmental Engineering Laboratory, Beijing Institute of Spacecraft Environment Engineering, Beijing 100094, China}

\author{Bin LIAO}
\affiliation{College of nuclear science and technology, Beijing Normal University,100875,Beijing, China}

\author{Lixiang JIANG}
\affiliation{Science and Technology on Reliability and Environmental Engineering Laboratory, Beijing Institute of Spacecraft Environment Engineering, Beijing 100094, China}

\author{Sidao NI}
\author{Tianfang LIU}
\affiliation{State Key Laboratory of Geodesy and Earth's Dynamics, Innovation for Precision Measurement Science and Technology, Chinese Academy of Sciences, Wuhan 430077, China}

\author{Zheng WANG}
\affiliation{School of Nuclear Science and Engineering, North China Electric Power University, Beijing 102206, China}

\begin{abstract}
  We present a cosmic ray muon tomographic experiment for measuring the muon flux and imaging the tunnel overburden structures in Changshu, China. The device used in this study is a tracking detector based on the plastic scintillator with SiPM technology, which can be conveniently operated in field works. The compact system with sensitive area of $6400 cm^2$ can measure the angular distribution of cosmic muons. It's able to image the overburden density length from the surface of overburden to the detector along the muon tracks. The open sky muon flux measurement outside the tunnel has a good agreement with the modified Gassier Formula model. The distributions of muon flux at three positions inside the tunnel are very similar to that of open sky. Assuming the average density of overburden compact sandstone is $2.65 g/cm^3$, the overburden thickness can be obtained from the density length derived from the difference of muon flux inside and outside the tunnel. Moreover, for known penetrated lengths (i.e., topography of overburden), the density anomalies of the overburden can also been obtained. This study suggests a potential application for imaging and detecting subsurface structures in civil engineering, tunnels or caverns with the cosmic ray muon telescope.
    \\
    \\
  $Keywords$: muon flux, plastic scintillator telescope, muon tomography, overburden structure imaging

\end{abstract}


\maketitle


\section{Introduction}

\label{sec:intro}
Muons are secondary particles of galactic cosmic rays. Naturally, galactic cosmic rays rain steadily from all directions to the Earth. The interaction of these high energy ions and the atmosphere nuclei produce a cascade of secondary particles. Benefit from the isotropic distribution, muons can be detected at anywhere on the earth in principle with a good approximation. The muon flux reaching the surface of the Earth is about $1 muons/min/cm^2$\cite{Hagiwara:2002}.  Muon can penetrate hundreds of meter rocks and can be used to imaging the subsurface structures of large targets because of the high energy and large mass properties \cite{Bogdanova:2006}. This is known as the muon tomography, which is an effective and powerful technique to image the density structures of geological bodies. This kind of measurement is based on the energy loss of high energy muons in matter. Since the muon flux is steeply decreased as a function of the energy, therefore passed matter above the detector modifies the threshold energy for the detectable particles and correspondingly the flux. Due to this fact, the measured muon flux correlates with the density length of traversed material, which is the key issue in the imaging of the internal anomaly structure of large objects\cite{Bryman:2014}\cite{Lesparre:2010}\cite{Tanaka:2001}\cite{Tanaka:2003}.

Subsurface structures of the Earth is one of the most important aim in geophysics. However, density structure is still a challenging problem in Geophysics.  With seismological and magnetotelluric measurements, only wavespeed and conductivity structures of the Earth can be obtained. Gavity data can provide constraints on the dnesity structure but with low resolution. Fortunately, Muon is sensitive to the subsurface density structures with high-resolution, which can be used to imaging the density structures reliably. Muon tomography technology has been used in geological studies since 1950 and was first used to explore the overburden density structure above a mountain tunnel \cite{Ceorge:1955}.  In 2017,the measurement performed by the frame of the METROPOLIS project obtained the evidence of the existence of unexpected cavities in Mt. Echia in Naples  \cite{Saracino:2017}. In the same year, Charlotte Rowe and collaborators recovered a 3D tomography of a tunnel beneath Los Alamos, New Mexico \cite{Guardincerri:2017}. The recent study performed by the same group illustrated the potential of inverting for shallow geologic structures by combining gravity and muon measurements \cite{Cosburn:2019}. Also, this method has been widely applied in the study of the internal structure of volcanoes,such as \cite{Beauducel:2008}\cite {Marteau:2012}\cite{Fehr:2012}\cite{Tanaka:2009a}\cite{Tanaka:2009b}

Muon tomography is also a very promising technique for subsurface resource exploration and monitoring. In order to facilitate the detection of heavy metal bodies such as uranium, gold or lead, muon measurement experiments have been performed on existing mines\cite{Bryman:2015}\cite{Baccani:2019}. Monitoring tunnel voids can be time-consuming and may involve many personnel working in potentially hazardous environments.  As a non-invasive technique, muon tomography can detecting overburden density anomalies reliablely and accurately\cite{Thompason:2019}.

For the future development of muon tomography in the subsurface resource exploration and monitoring applications, we conducted a new muon tomography experiment in a tunnel in Changshu, China. This study introduces the basic structure of the muon telescope, a portable and durable tracking-detector based on plastic scintillator. We described the approach of muon flux measurement and the overburden structure imaging. The detector was placed at three different positions inside the tunnel and one position outside the tunnel, and all the muon flux  measurements results at each point are given. Based on the difference of muon flux between inside and outside the tunnel (i.e. muon flux ratio), we get the overburden structures above the three points inside the tunnel. The experiments well recover the overburden structures with this non-invasive technique, and can obtain the density structures with known overburden thickness distribution, or obtain the overburden thickness with known density distribution.

\section{The Experiment Site and Muon Telescope}
\label{sec:exptel}
\begin{figure}
  \includegraphics[width=7cm]{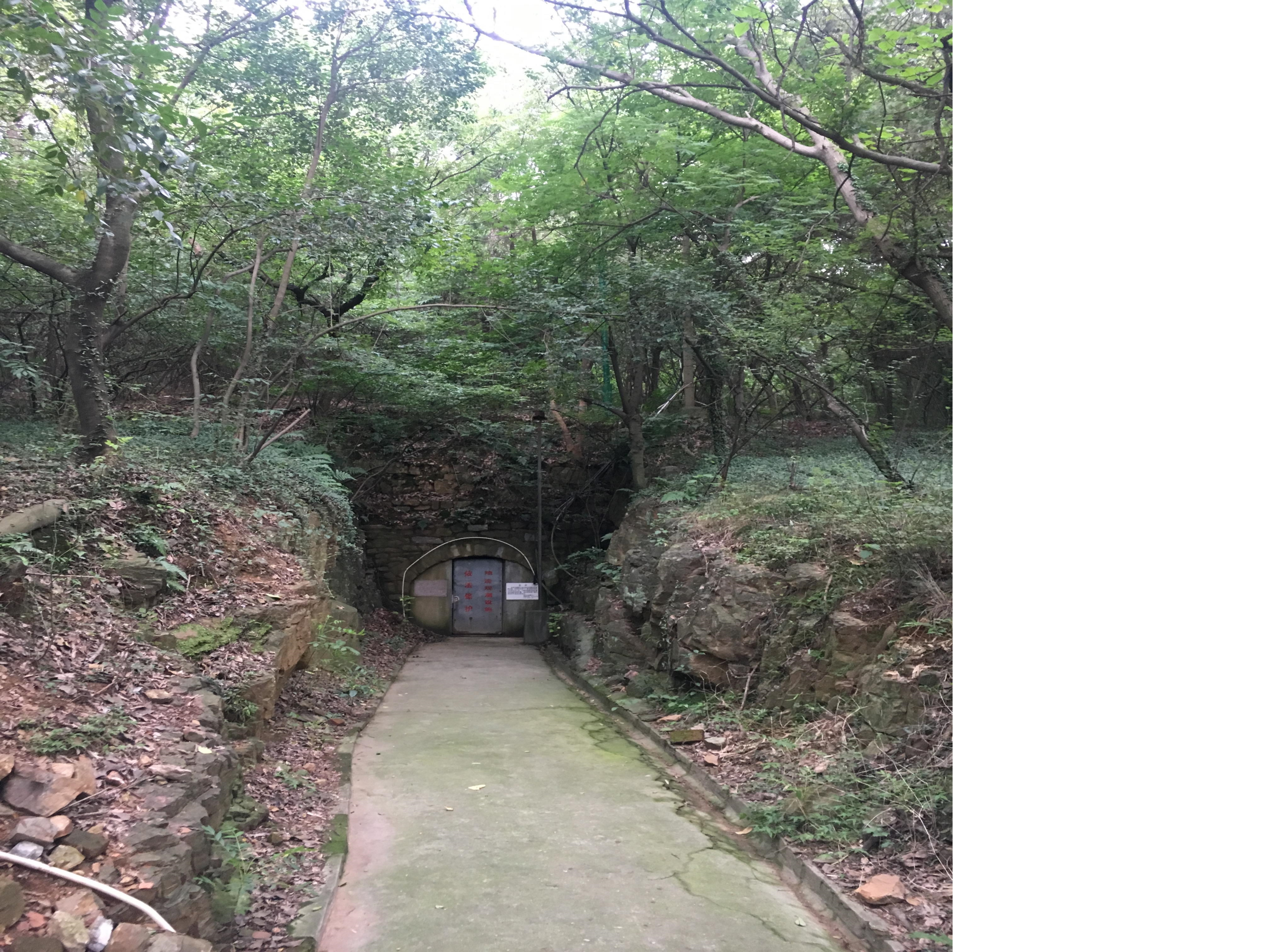}\\
  \caption{The picture of the tunnel entrance.}\label{p:entrsrtm}
\end{figure}

The tunnel experiment is located in the northwest of Changshu, Jiangsu Province. The tunnel has a length of about $250m$ , a height of $2.35m$ and a width of $2.5m$. The temperature inside of the tunnel is around $17.3 $ degrees Celsius all year round. The tunnel bedrock is compact quartz sandstone with the average density of $2.65~g\cdot cm^{-3}$, which is a dense and hard rock. The average overburden thickness is around $20m$, and ranges from $5m$(at entrance) to $40m$.  The picture of tunnel entrance are shown in Figure. \ref{p:entrsrtm}.

\begin{figure}[htbp]
\centering
\begin{minipage}[t]{0.5\textwidth}
\centering
\includegraphics[width=6.5cm]{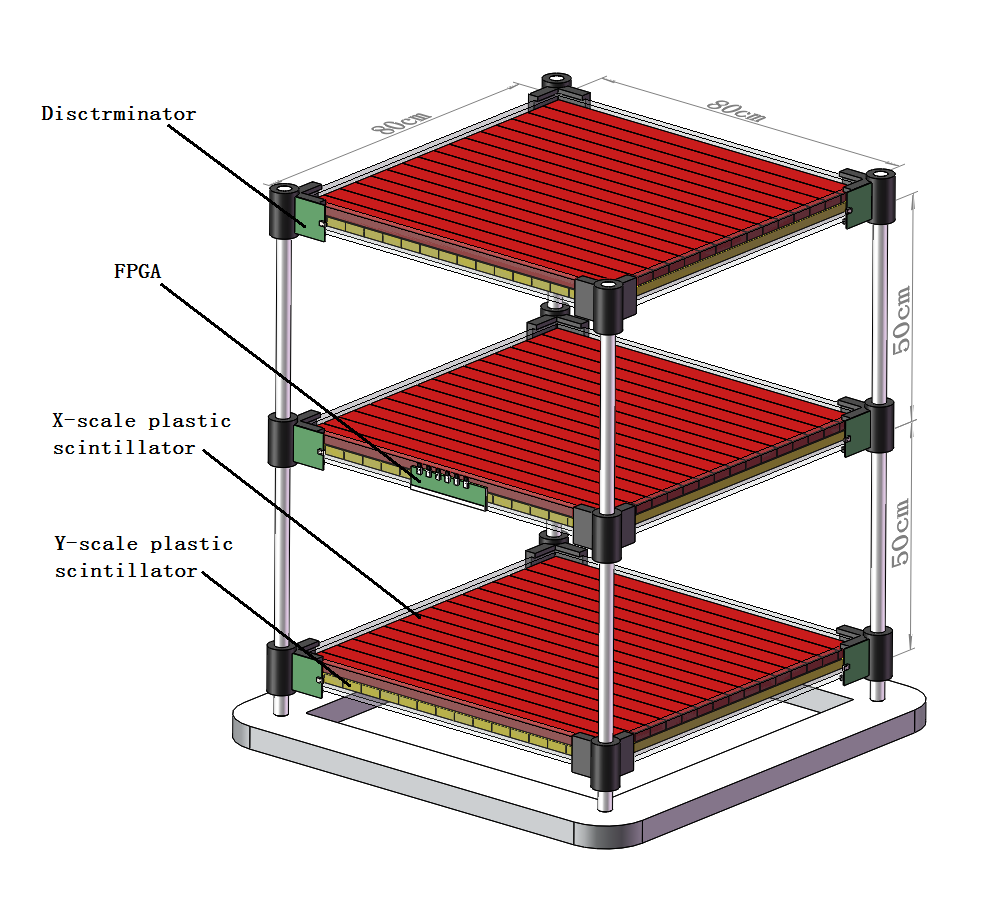}
\end{minipage}
\begin{minipage}[t]{0.48\textwidth}
\centering
\includegraphics[width=5.5cm]{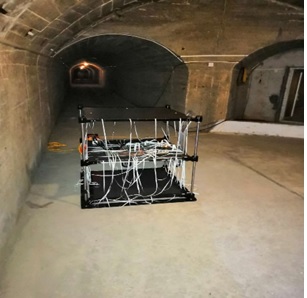}

\end{minipage}
\caption{Overview of the muon telescope.The left plot is the schematic structure and the right plot is the detector at tunnel.}\label{p:muondetector}
\end{figure}

The muon telescope is a three-layer plastic scintillator detector as shown in Figure. \ref{p:muondetector}. We designed it in a light-weighted and portable way to use it in outdoor situations. Each layer contains two planes, each plane is made up of 16 plastic scintillator strips, so there are 96 strips in total. The two planes in each layer is placed orthogonally, to provide $x$ and $y$ coordinates.

The size of the plastic scintillator strip is $80cm \times 5cm \times 1cm$, which makes the total effective area is $80cm \times 80cm$. Each plastic scintillator is covered by highly-reflective enhanced specular reflector (ESR) to improve the scintillation light collection and two silicon photo-multiplier (SiPM) at the end of each strip to collect the light.
The distance between each layer can be adjusted and the maximum distance between the top and bottom layer is $100cm$. For the measurement at Changshu tunnel, the distance between the each layer is set to be $50cm$.

The model of plastic scintillator is HND-S2 made by Chinese factory, which has relatively high light yield, $50\%-60\%$ compared with paranaphthalene. The wavelength of the emission scintillation photon is typically in the $395 nm$ to $425 nm$ range. The attenuation length is larger than $2m$ and the decay time is around $2.8 ns$ and the refractive index of this scintillator is $1.58$, which makes it totally compatible with regular SiPM. Other character such as the H/C ratio is close to $1.1$ , the density is $1.05 g\cdot cm^{-3} $  and the softening temperature is $348K - 353K$.  The SiPM used is MicroFC-30035-SMT provided by SensL company. The wavelength at the highest quantum efficiency is $420 nm$, which matches the emission spectral of the plastic scintillator. It also has other nice features, such as low operating voltage, excellent temperature stability, robustness, compactness, output uniformity, etc.

Data from each strip is processed using the front-end readout electronics (FEE). FEE is equipped with 6 discriminators and 1 FPGA. The discriminator is triggered logical OR signals from the two SiPMs at the ends of strips. After the signal was trigged, the discriminator amplifies the signal and if it exceeds the $32mV$ threshold, the strip will be considered as fired one, we keep the event with more than 3 strips fired from different layers. The 32 channel signals of each layer array plane are processed by one discriminator. FPGA receives $32\times 6=192$ digital trigger signals from 6 discriminators and the data is transmitted to a computer via a mini-USB cable.

\section{Muon Flux Measurement}
\label {sec:muonflux}
The muon detector has been deployed from April 22 to July 8, 2019 at the tunnel. Apart from the handing and debugging time for the detector, the tunnel data-taking campaign consisted of up to 73 days. The system was operated continually in the site for everyday. Before operation in the tunnel one open-sky measurement was taken for later detector calibration. Data taking was performed in runs of about $1$ hour with at least 3 strips fired, storing all the buffer events.  We successfully took $71$ days data in the tunnel and $2$ days data at the entrance of the tunnel for the open sky condition. Approximatively $3.01\times 10^{7}$ events were obtained in the tunnel and $9.06\times 10^{6}$ events outside the tunnel. Table. \ref{datataking} presents the data acquired at each of the detector positions. In this section, the analysis procedure to get muon flux at each point is introduced.

\begin{table}
\centering
\caption{\label{datataking} Muon data acquisition were deployed inside the tunnel (Position No.1, 2 and 3) and outside the tunnel (open sky). $D_1$ denotes the distance between the location and entrance gate. Total events is all the events kept with 3 strips fired, 6 layers liner events are selected liner events from total event pool with coincident hits in all six layers as descried in the below section.}
\smallskip
\begin{tabular}{cccccc}
\hline
Position No. ~~& ~~$D_1(m)$~~& ~~acquisition time (s) ~ & ~~total events per day~~ &~~ 6 layers liner events per day  \\
\hline
1 & 246 & ~ 164,1600 & ~$ ~5.32 \times 10^{5} ~ $ & $~ 1.12 \times 10^5 ~$ \\
2 & 231 &  ~691,200~ & $~4.35 \times 10^{5} ~$ & $~ 0.89\times 10^5 ~$ \\
3 & 200 & ~380,1600~ & $~3.75 \times 10^{5} ~$ & $~ 0.73\times 10^5 ~ $\\
~~~Open sky~~~ &~~~0 ~~~&  ~~~73,101 ~~~& $~4.53 \times 10^{6} ~ $ & $~ 9.80\times10^5 ~$\\
\hline
\end{tabular}
\end{table}

\subsection{Data Analysis}
\label {subsec:dataana}
The data analysis starts with the reconstruction of the muon tracks event by event. Muon track can be reconstructed in cartesian coordinates $(x, y, z)$ in the detector system, where $x$ and $y$ represent the coordinates in the detection plane while $z$ axis is perpendicular to the detection plane.  In the following analysis, the muon track deflection angle caused by scattering is ignored. The track reconstruction uses only events with coincident hits in all six layers of the detector, and keep events with only one hit per layer.

The reconstruction algorithm preselects coincident hits in a $20 ns$ time window in all six detection layers, from which tracks are obtained analytically through a straight line procedure. The muon track reconstructed by six layers is much cleaner which could eliminate the fake muon event of other particles or electronic noise striking on two matrices at the same time. In this sample, we select events containing just a single muon track which will lose some of real muon events, but we kept all those marked as ``golden track".

\begin{figure}
 \centering
  \includegraphics[width=.65\textwidth]{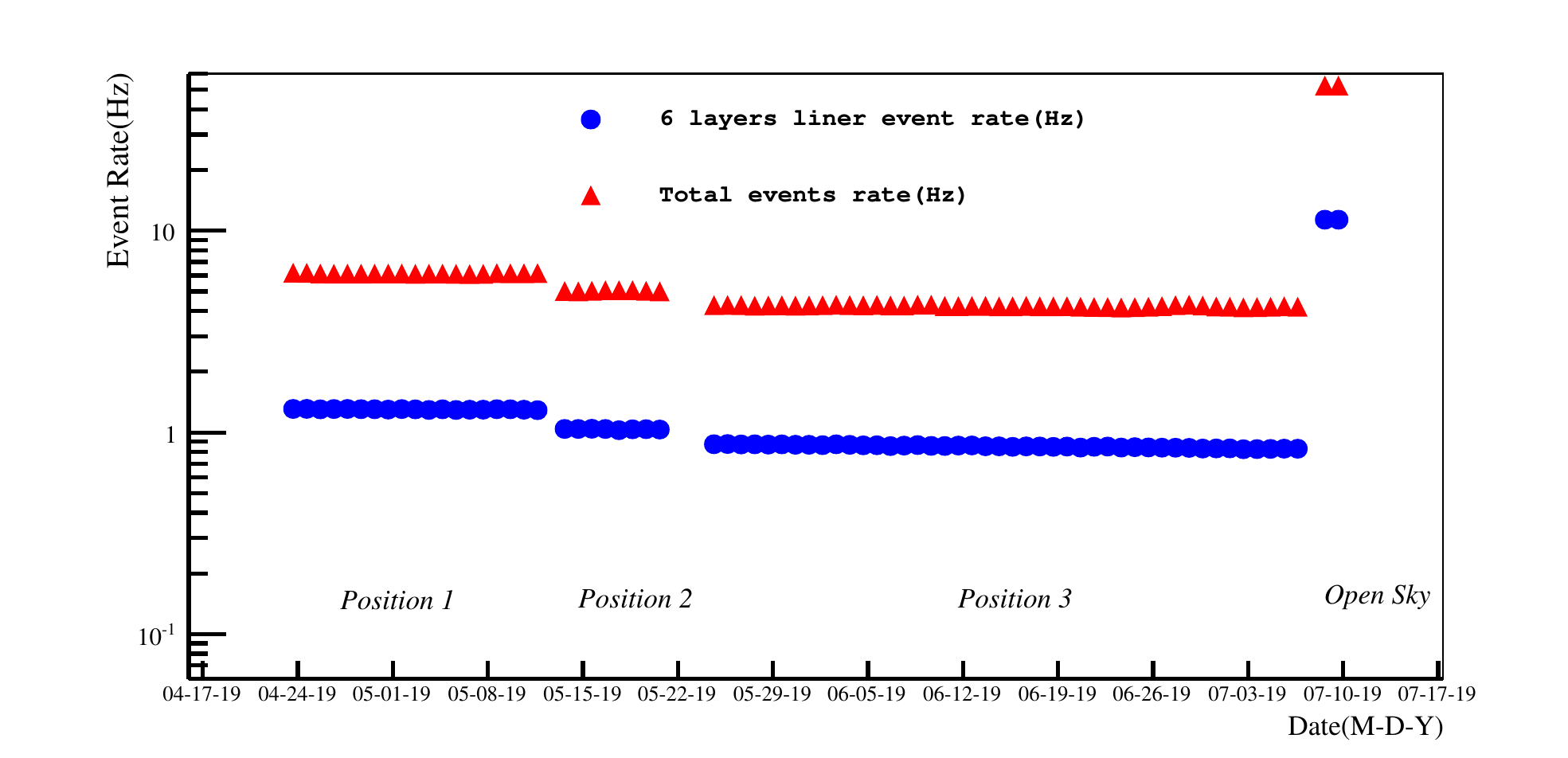}\\
  \caption{The muon event rate of golden track(6 layers liner events) and total collected events at each location.The data from 24th April to 12th May are collected at the first point of inside the tunnel, from 14th May to 21th May are belong to the second point of inside the tunnel, from 25th May to 7th July are the third point of inside the tunnel, and the last two days, 9th and 10th July are the open sky data. The red triangle is the events rate of the total collected data, and the blue circle is the events rate after 6 layer linear selected data.} \label{p:rate}
\end{figure}

\begin{figure}
 \centering
  \includegraphics[width=.65\textwidth]{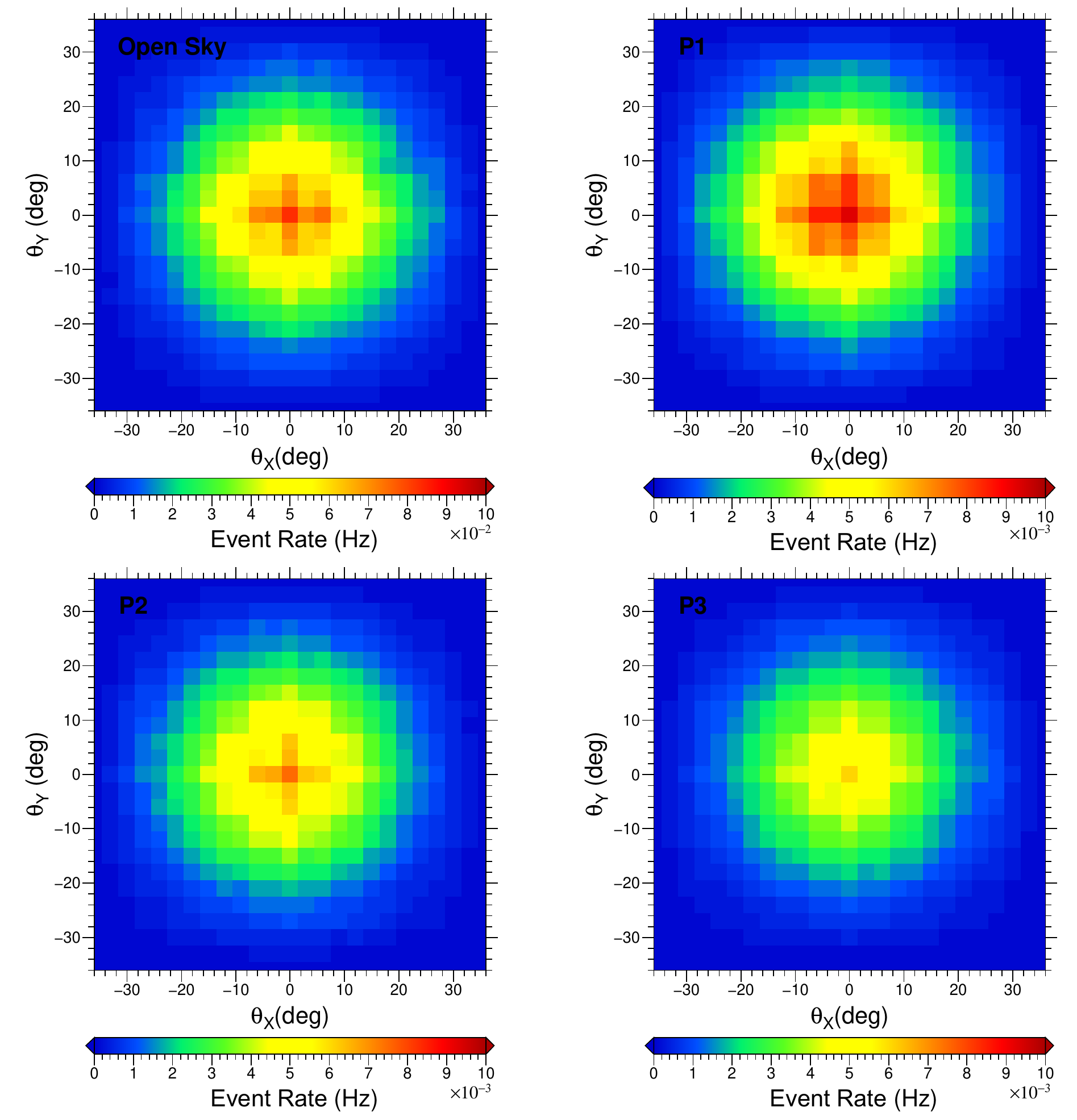}\\
  \caption{The distribution of $(\theta_x,\theta_y)$ of reconstructed golden muon event rate at open sky and position1,2, 3 inside the tunnel. } \label{p:theta}
\end{figure}

We define the pre-selection efficiency as the number of events that satisfy the straight line procedure divided by the number of events with 3 strips fired. At the end of this selection, we estimate an average data selection efficiency around $21.0\%$.  During the whole data taking period to the open sky, the golden track rate is $11.34 Hz$, for the other 3 points inside the tunnel, the rate as follows: $1.30 Hz$, $1.04 Hz$, $0.85 Hz$, respectively. The correspond golden events number are listed in the last column of Table.\ref{datataking}. Figure.\ref{p:rate} gives the map of muon event rate of golden track measured at each location. The muon event rate of selected golden track with $\theta_x$ and $\theta_y$ distribution as Figure.\ref{p:theta}. $\theta_x$ or $\theta_y$ denotes projection of a golden muon track zenith angle in the $x-z$ or $y-z$ telescope plane.

From the selected golden muon track to muon flux measurement, the detector performance and geometrical factor still need to be considered.  A useful quantity for characterizing the behavior of a detector meant for measuring muon flux is the effective surface, $S_{eff}$, which takes into account the detection performance and the geometrical acceptance of the detection setup. The geometrical acceptance $A(\theta_x,\theta_y)$ includes the effective area of the detector, solid angle and the geometrical factor. The detection performance $\epsilon(\theta_x,\theta_y)$ includes the detector efficiency and selection efficiency. Having the number $N_g$ of golden muons detected for each direction $(\theta_x,\theta_y)$ during a period $\Delta T$ , the integrated muon flux $I$ will be Equation.\ref{f:Muflux} as

\begin{equation}
I(\theta_x,\theta_y) = \frac{N_g}{S_{eff}(\theta_x,\theta_y)\times\Delta T}(sr^{-1} \cdot cm^{-2}\cdot s^{-1} )\label{f:Muflux}
\end{equation}

where

\begin{equation}
S_{eff}(\theta_x,\theta_y) =  A(\theta_x,\theta_y) \times \epsilon(\theta_x,\theta_y)\label{f:Seff}
\end{equation}

The use of detector matrices formed with a limited number of pixels of finite size causes geometrical acceptance $A(\theta_x,\theta_y)$. The acceptance of the telescope is a key parameter since the goal of the project is to assign an opacity and therefore a density to the target from an absolute measurement of flux. The acceptance of the telescope includes the area of the detector plane, solid angle of each pixel $\Omega(\theta_x,\theta_y)$ and the geometrical factor $\Lambda(\theta_x,\theta_y)$.

 The quantization of solid angel $\Omega(\theta_x,\theta_y)$ depends on the number of pixels $N_x\times N_y$, on their size d, and on the distance D of separating the two detector matrices. In our case, $N_x=N_y=16$,$d=5cm$ and $D=100cm$. Typical solid angel curves in this experiment are shown in the left of Figure.\ref{surface}. Taken as an angular resolution, solid angle can privde the spatial resolution of one telescope. For example, with $D = 100 cm$ and for the $(0,0)$ vertical direction, the solid angle is about $2.5\times10^{-3} sr^{-1}$, this allowing to detect heterogeneities with a size of about 10cm at a distance of 5m, also means the spatial resolution of our telescope can reach 10cm at 5m distance to the observation.

The geometrical factor just in the case of the muon telescopes made of parallel layers of finite dimensions, the particles are
required to physically cross the outer layers in order to induce coincident signals. This factor can be analytically calculated for a perfect
telescope with square layers of size $d^2$ and a distance $D$ between the two outermost layers. In the right of Figure.\ref{surface}, we show the geometrical acceptance of the detector for each detection element. From this figure we can see that the largest effective area corresponds to the vertical area for every detection element which can be as large as $\simeq 16 cm^{2}\cdot sr^{-1}$. However, the smallest effective areas  corresponds to the body diagonal directions.
For large $(\theta_x,\theta_y)$, $ A(\theta_x, \theta_y)$ is close to $0$, which means only a few muon events can be recorded in these regions. In other words, the effective detection region lie in the range between $-36$ deg and $36$ deg in the detector frame.

\begin{figure}[htbp]
 \centering
  \includegraphics[width=.85\textwidth]{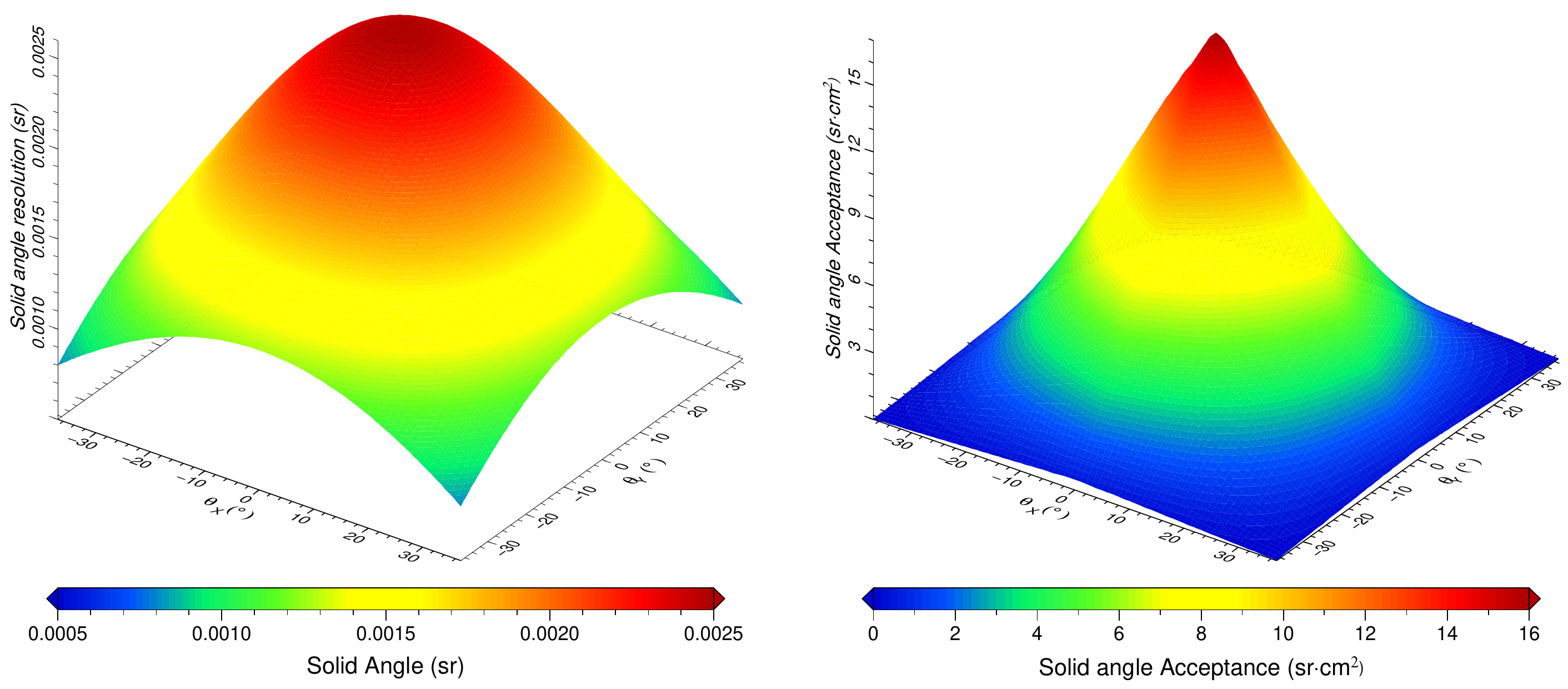}\\
\caption{The left plot is the angular resolution of the telescope equipped with two $16 \times 16$ matrices with pixel size $d = 5cm $ and separated by
$D=100cm$.The right plot is geometry acceptance of the telescope.} \label{surface}%
\label{surface}
\end{figure}

\begin{figure}
 \centering
  \includegraphics[width=.5\textwidth]{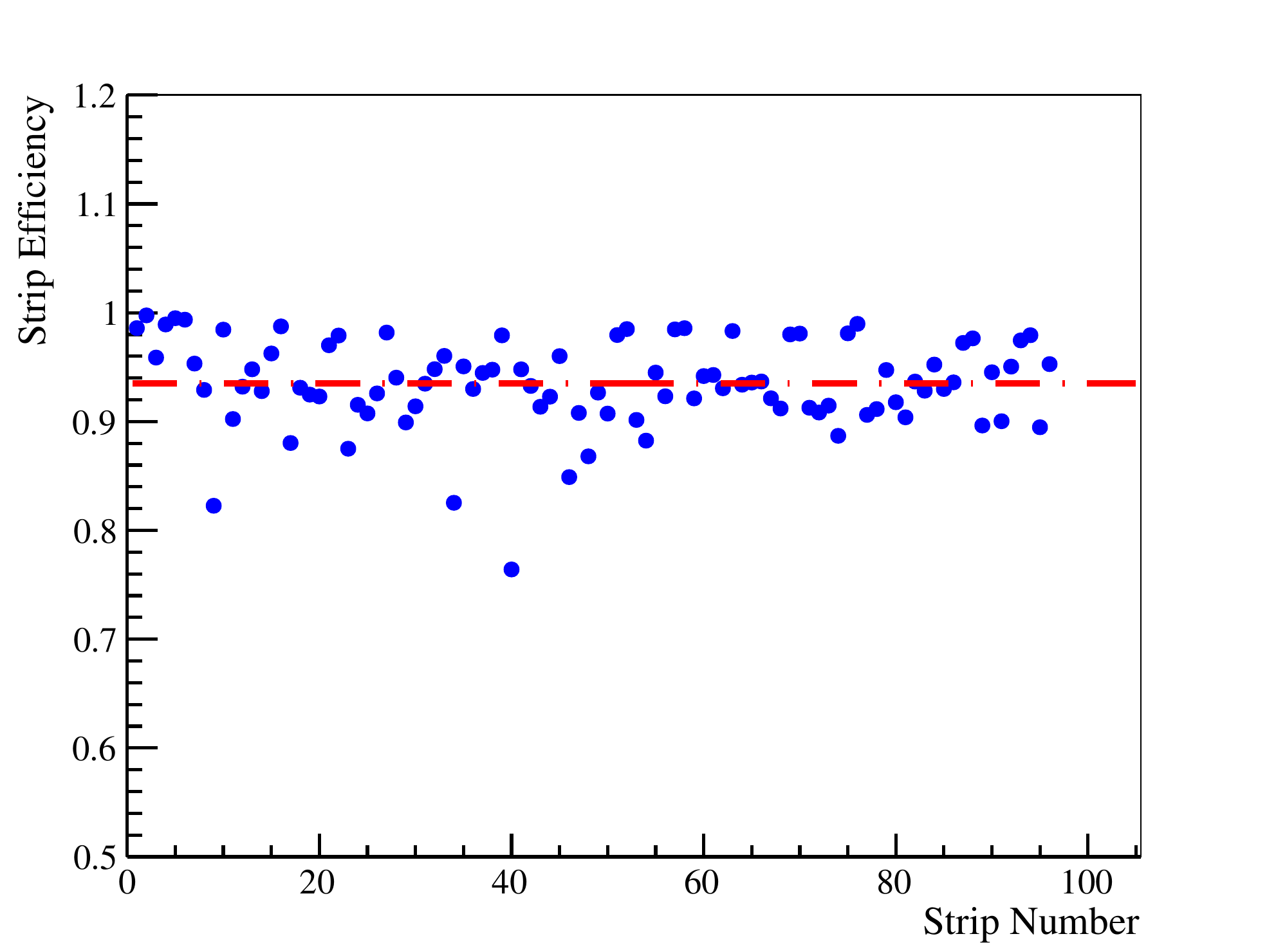}\\
  \caption{The efficiency of each strip in the telescope, fitted by the linear function as red line show, gives the average efficiency is 0.94. } \label{detectoreff}
\end{figure}

The detection performance $\epsilon(\theta_x,\theta_y)$ includes the detector efficiency and selection efficiency.
The detector efficiency $D_{eff}$ calculated strip by strip. Only the vertical events were used to do the detector efficiency calibration. Figure.\ref{detectoreff} are the efficiency with the strip number, except the 40th strip with a little bit low efficiency, all the others are good enough and the average efficiency is around 93.5\%. The final straight line efficiency will be get from all the linear six hits, this gives the average detection efficiency is around $64.3\%$ for the muon track. About the selection efficiency, benefit from that the detector are stable in time based on scintillator, an average trigger efficiency of $80\%$ is estimated.

\subsection{Muon Flux Measurement Results }
\label {subsec:fluxresults}
\begin{figure}
  \centering
  \includegraphics[width=.65\textwidth]{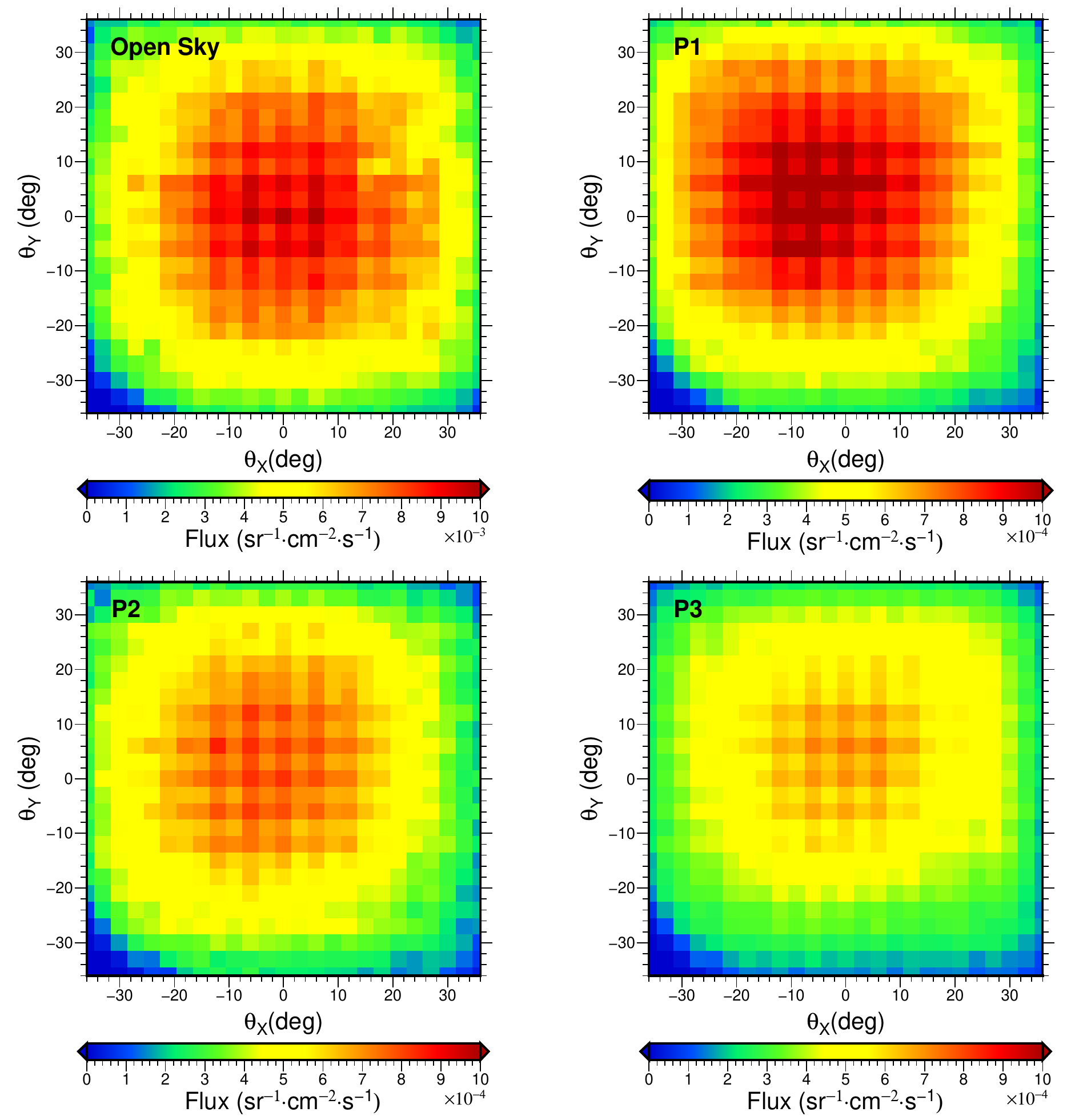}\\
  \caption{The distribution of $(\theta_x,\theta_y)$ of measured muon flux at open sky, position1,2 and 3 inside the tunnel. The measured muon flux are getting from the golden muon event rate in Figure.\ref{p:theta} by divided the detector performance and geometrical factor described in paper.}\label{p:muonflux}
\end{figure}

 Based on Equation.\ref{f:Muflux}, as a function of $(\theta_x,\theta_y)$, the muon flux is estimated by dividing the number of golden tracks in each angular bin to $S_{eff}(\theta_x,\theta_y)$ and a period $\Delta T$ in Table.\ref{datataking}. The measured muon flux, averaged over bins in the control regions is represented as a function of the $(\theta_x,\theta_y)$ in Figure. \ref{p:muonflux}.

The muon flux in vertical direction($\theta_x=\theta_y=0$ direction, i.e. Zenith angle $\theta=0$) of four positions are shown as below:\\
  $I_{(0,0)}=9.98\times 10^{-3}cm^{-2}\cdot s^{-1} \cdot sr^{-1}$  (Open sky)\\
  $I_{(0,0)}=1.03\times 10^{-3}cm^{-2}\cdot s^{-1} \cdot sr^{-1}$  (P1)\\
  $I_{(0,0)}=0.83\times 10^{-3}cm^{-2}\cdot s^{-1} \cdot sr^{-1}$  (P2)\\
  $I_{(0,0)}=0.71\times 10^{-3}cm^{-2}\cdot s^{-1} \cdot sr^{-1}$  (P3)\\

\begin{figure}
 \centering
  \includegraphics[width=.5\textwidth]{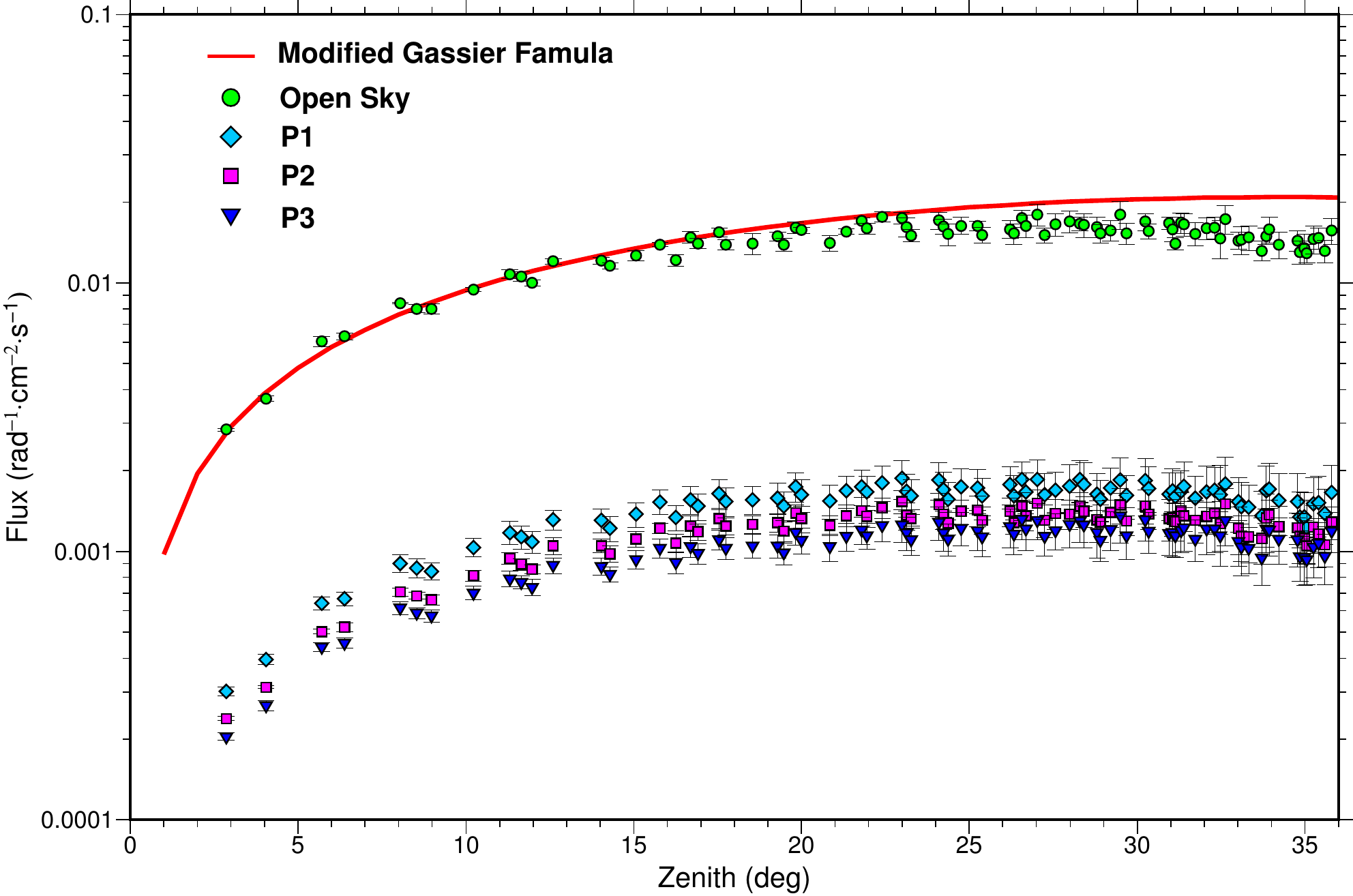}\\
  \caption{Muon flux distribution over the zenith angle $\theta$ at open sky and position 1,2,3 inside the tunnel. The bar in each point gives the range of muon flux with the same zenith angle $\theta$ but different azimuth angle $\phi$. From the figure one can see that the open sky data is coincidence with modified Gassier Formula model, and the distributions of the flux with overburden are very similar to the distribution at the open sky data.} \label{p:muflxtheta}
\end{figure}

  Muon flux distribution over the zenith angle $\theta$ is also taken into account at each point, dependency on azimuth angle $\phi$ being uniform. Results of our experiment are shown in Figure.\ref{p:muflxtheta}. It is seen from Figure.\ref{p:muflxtheta} that the angular distributions is satisfactorily coincident with the modified Gassier Formula \cite{Guan:2015}. The bar in each point gives the range of muon flux with the same zenith angle $\theta$ but different azimuth angle $\phi$. Zenith angle $\theta$ dependency at different overburden structures, which is of interest by itself\cite{Bogdanova:2006}. As it is seen from results, the distribution of position 1, position2 and position 3 turn out to be very similar to the distribution at the open sky data only with less muon flux number, contrary to the frequent expectation that they should become narrower, since the absorption length increases with angle.

\section{Overburden Structure Results}
\label{sec:overstru}
 The main aim of our measurements in the tunnel system was to verify the applicability of the muon telescope for underground structure, i.e., the surface reconstruction and/or estimate rock inhomogeneity. Here we investigated an underground structure method using cosmic muons. In this section, we will show the approach how to get the overburden density length structure,$\rho\cdot L$ from cosmic muon measurement. In the three measurements, the detector is located at the same elevation inside the tunnel. It is feasible to estimate the average density of the overburden in different directions along a specific line of sight of the detector with known overburden thickness. Moreover, if the average density is known, the overburden thickness distribution can also be obtained.

 The relation between the muon flux and density length is derived from the theory equation firstly. By comparing the experiment results with this relation,the corresponding density length in different zenith and azimuth angle for each experiment site can be calculated. Here we introduce the method to get the relation between muon flux and overburden thickness.
 The number of muons expected to be acquired inside the tunnel $N(\theta)$ can be written as Equation.(\ref{f:muonfluxdef})

\begin{align}
  N_{in}(\theta) =  \int^\infty_{E_{min}}\Phi(\theta, E)  dE \cdot S_{eff} \cdot \Delta T_{in}, \label{f:muonfluxdef}
\end{align}
where $\Phi(\theta, E)$ is the differential muon flux as a function of the zenith angle $\theta$ and the muon energy $E$, this paper using modified Gaisser Formula \cite{Guan:2015} model, $E_{min}$ is the minimal kinetic energy that a muon must possess to reach the detector location without being absorbed by the overburden. $\Delta T_{in}$ is the data taking time and  $S_{eff}$ is the calibration factors as described in the above section. By comparing the measured muon flux at different positions, one can get the $E_{min}$ at each point. Using the relation between $E_{min}$ and density length in Table.EIV-6 of Ref. \cite{Groom:2001} to get the density length of the overburden.

\begin{figure}[ht]
  \centering
  \includegraphics[width=.5\textwidth]{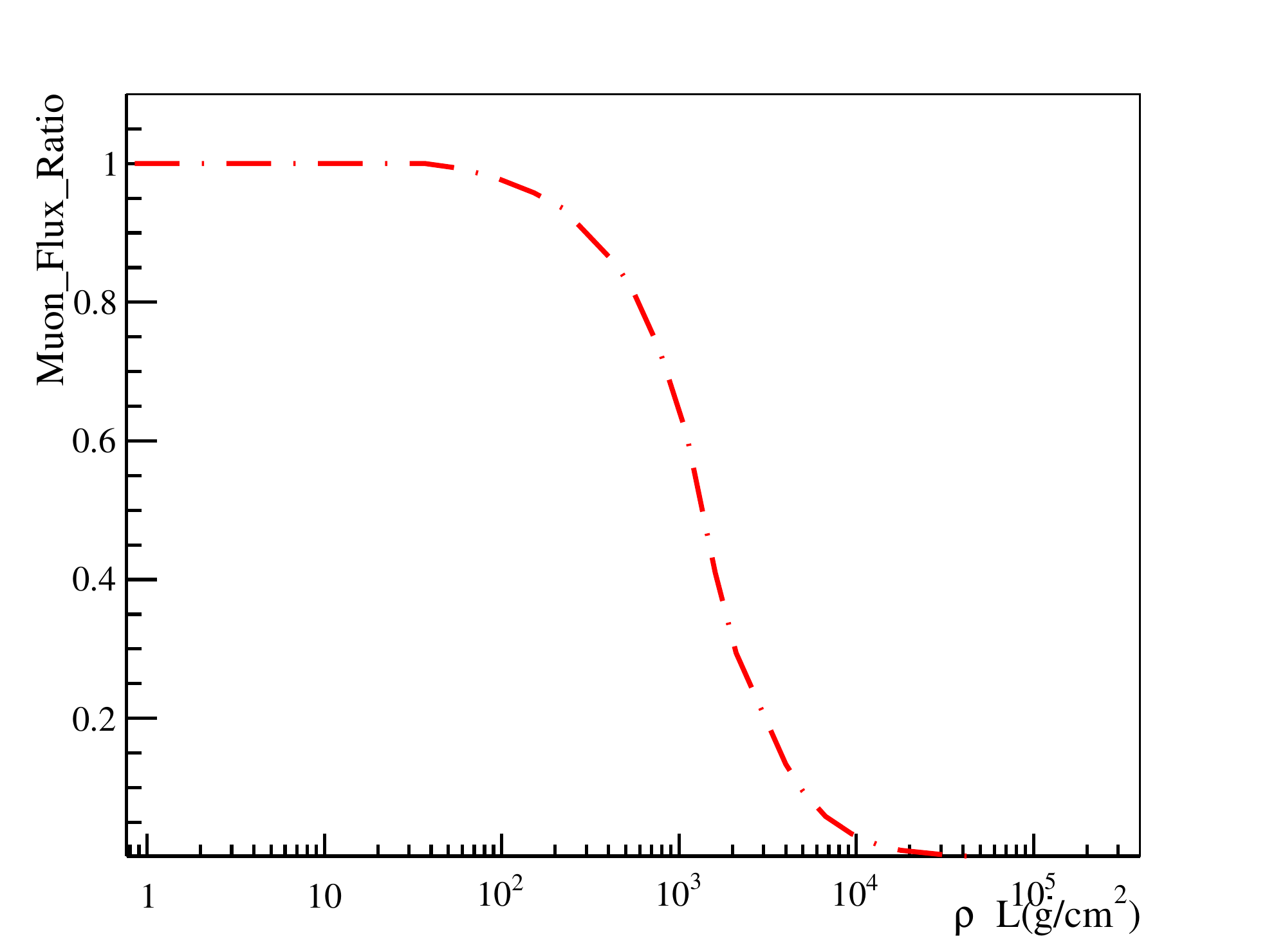}\\
  \caption{The muon flux ratio with the density length.}
  \label{p:fluxratiothickness}
 \end{figure}

Usually, it is difficult to get precise measurement muon flux, since from measurement muon number to muon flux, we need to take care calibration factors $S_{eff}$ very carefully. To eliminate the detector's performance influence, the muon flux ratio will be often used in the relation derivation. This ratio $K^a$ is defined as the muon flux in the tunnel to the muon flux in the open sky as,
\begin{align}
    K^a=\frac{N^{\nu}_{in}(\theta)/\Delta T^{\nu}_{in}}{N^{\nu}_{out}(\theta)/\Delta T^{\nu}_{out}}, \label{f:attenuation}
\end{align}
where subscript $in, out$ denote data selected inside the tunnel and outside the tunnel(the open sky data). From Equation.\ref{f:attenuation} we can see that the $S_{eff}$ is canceled, only need to normalized to the data taking time. The muon flux ratio with different density length \cite{Groom:2001} are plotted in Figure.\ref{p:fluxratiothickness}.

\begin{figure}[ht]
  \centering
  \includegraphics[width=.65\textwidth]{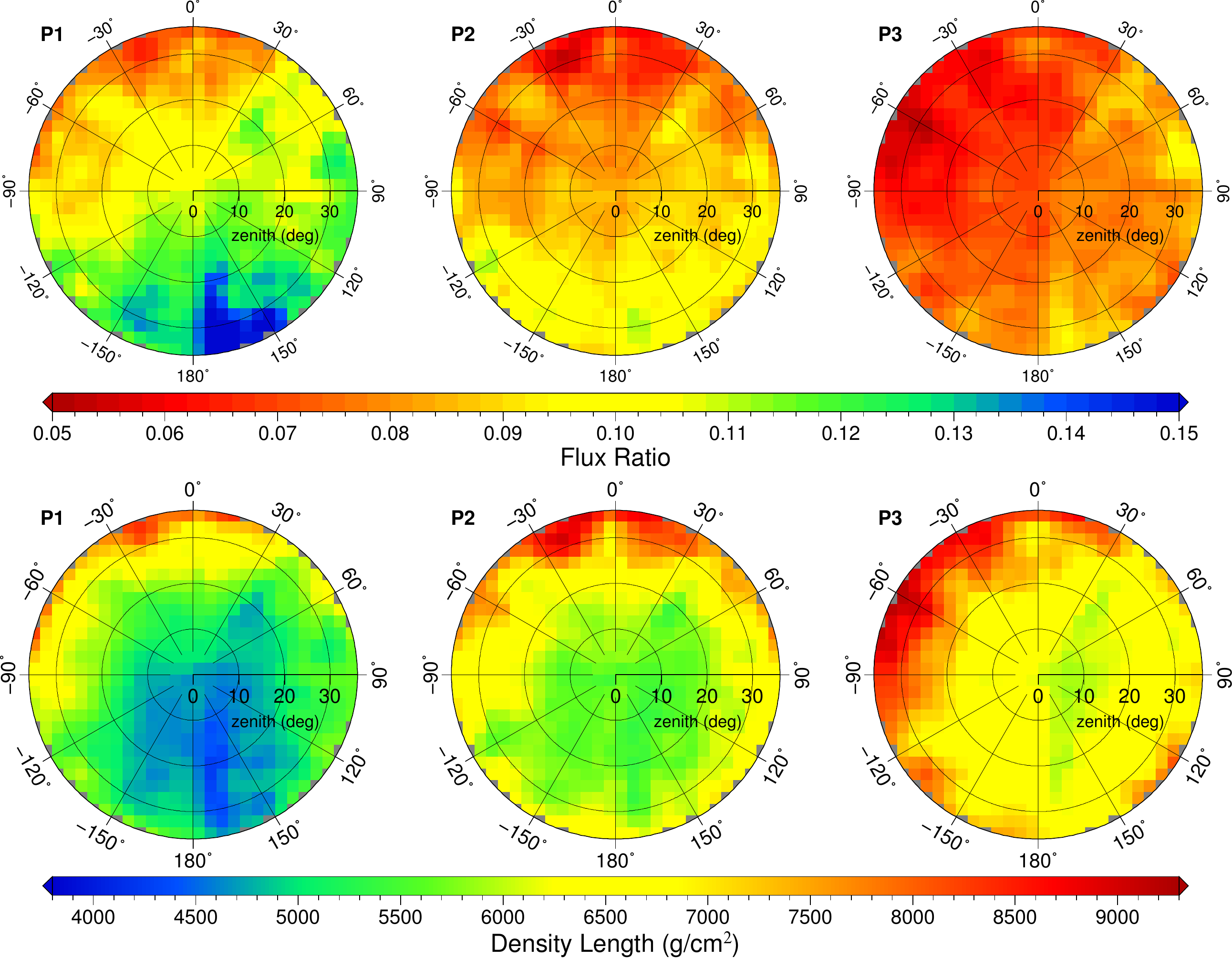}\\
  \caption{The top plot is the measured muon flux ratio at 3 points in the tunnel with zenith angle $\theta$ and azimuth angle $\phi$. The bottom is the overburden density length distribution at 3 points with different of $\theta$ and $\phi$ along a specific line of sight of the detector. }
  \label{p:thicknessfluxthphi}
 \end{figure}

 Compare the measurement $K^a$ in our experiment and Figure.\ref{p:fluxratiothickness}, the measurement density length results of each points at the tunnel are shown in Figure. \ref{p:thicknessfluxthphi} in detector system.

\begin{figure}[ht]
  \centering
  \includegraphics[width=.55\textwidth]{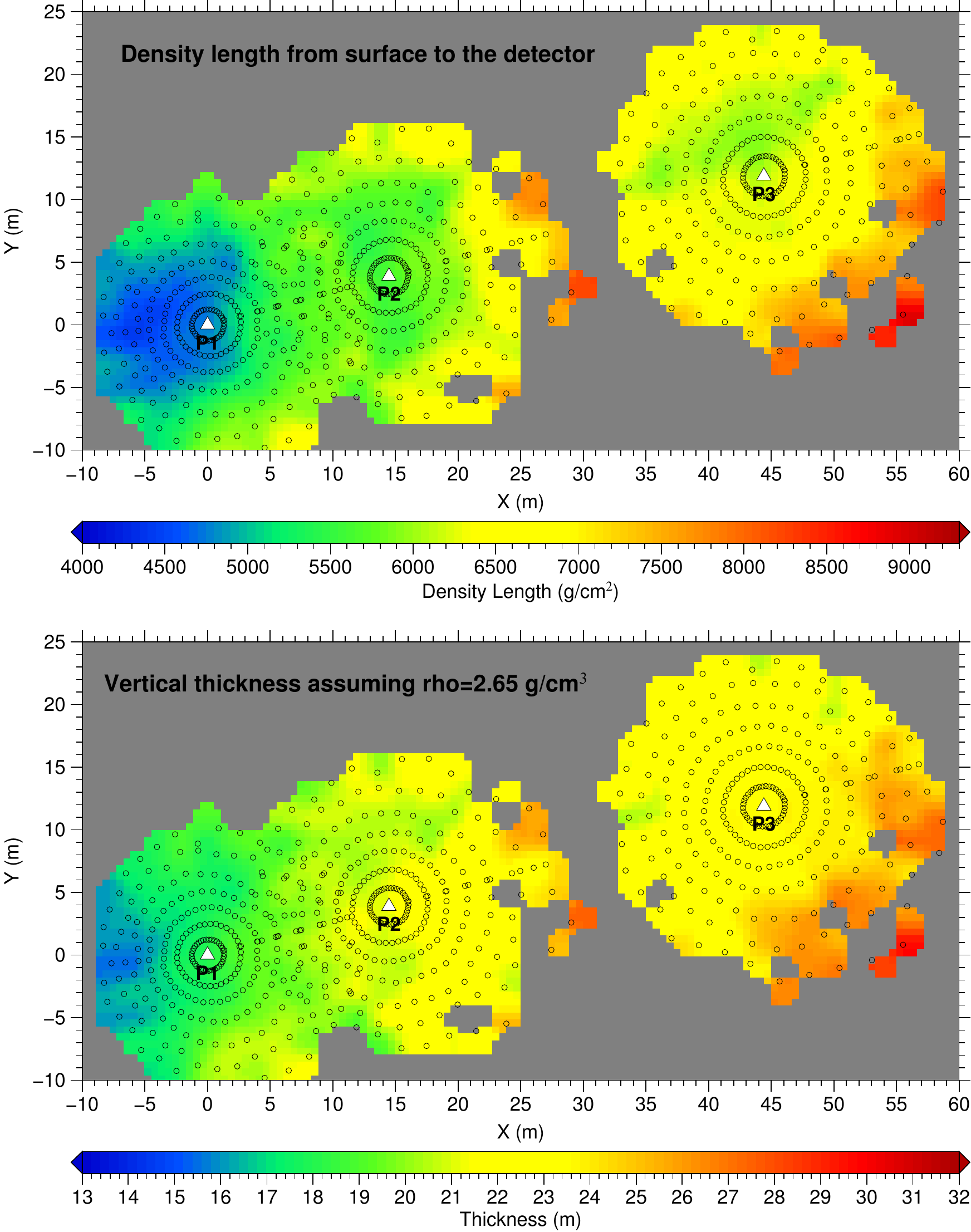}\\
  \caption{The top is the overburden density length distribution at 3 points inside the tunnel with $(X,Y,Z)$ in geodetic coordinate system. The bottom is the map of overburden thickness distribution assuming the average rock density is $2.65g/cm^3$. }
  \label{p:thicknessxyz}
 \end{figure}

We expressed the same density length distribution but with $(X,Y,Z)$ in geodetic coordinate system as the top of Figure.\ref{p:thicknessxyz}. With the known topography of the tunnel, one can get the density distribution from the density length. In turn assuming an average rock density, the thickness of the overburden will be got. The bottom of the Figure.\ref{p:thicknessxyz} show the thickness distribution assuming the average rock density is $2.65g/cm^3$. From the figure we can see that the overburden thickness above each position does not change rapidly. The range of the overburden thickness above three positions are around $20.00 \pm 2.95 m$, $23.03\pm 2.75 m$, $25.18 \pm 4.44 m$ respectively. The standard derivation shows the variation range of the measured overburden thickness in different zenith and azimuth angles. Our results are consistent with the mountain topography above the tunnel.

\section{Conclusion}
 \label{sec:summary}

The cosmic ray muon experiments inside and outside the tunnel have been conducted to imaging the overburden structure. Cosmic ray muon tomography has been proved to be a feasible and powerful technique for the identification of void space and significant density anomalies of subsurface structures. Three measurements have been performed inside the tunnel of Changshu area with overburden thickness of $15\sim30$ meter rock equivalent. Muon tomography can also be performed at larger depths deep to a few hundred meters, at the cost of increased measurement time due to the limited acceptance and strong absorption of muon. The current muon telescope is based on scintillator and SiPM technology with reliable tracking performance and low power consumption. The study demonstrates the feasibility for subsurface structure imaging with cosmic ray muon. The muon tomographic approach is promising in underground tombs investigation, subway tunnels engineering, geophysical, or industrial applications. With finer segmented scintillator detector, and observations from different positions and angles around the study target, a tomographic inversion with higher resolution can be conducted to obtain the three-dimensional density structures.

\acknowledgments

This work is supported by National Natural Science Foundation of China(Grant No. U1865206), State Key Laboratory of Geodesy and Earth's Dynamics and National Key Laboratory Foundation (Grant No. 6142004180203).


\end{CJK*}

\end{document}